\documentclass[10pt]{article}

\usepackage{amsmath}
\usepackage{amsthm}
\usepackage{amssymb}
\usepackage{amscd}
\usepackage{epsf}
\usepackage{rotating}
\usepackage{wrapfig}

\chardef\bslash=`\\ 

\makeatletter
\def\verbatim{\interlinepenalty\@M \@verbatim
  \leftskip\@totalleftmargin\advance\leftskip2pc
  \frenchspacing\@vobeyspaces \@xverbatim}
\makeatother
\theoremstyle{plain}

\theoremstyle{remark}
\newtheorem{rem}{Remark}

\numberwithin{equation}{section}

\def\1I{\relax{\rm 1\kern-.25em \rm l}} 

\def\Rahmen#1#2#3 {
   \vbox{\hrule height#2
         \hbox{
               \vrule width#2
               \hskip#1
               \vbox{
                     \vskip#1{}
                     \hbox{#3}
                     \vskip#1
                    }%
               \hskip#1
               \vrule width#2
              }
         \hrule height#2
        }}
\def\href#1#2{#2}

\begin{document}

\thispagestyle{empty}
\rightline{hep-th/0404191}
\rightline{HU-EP-0423}
\vspace{2truecm}

\begin{center}

{\bf \Large  Killing Spinor Equations from Nonlinear Realisations} 
\end{center}

\vspace{.5truecm}

\newcounter{Institut}
\vspace{1.5truecm}
\centerline{\bf Andr\'e
  Miemiec\refstepcounter{Institut}\label{Inst_Andre}{}$^{*_{\theInstitut}}$
  and Igor Schnakenburg\refstepcounter{Institut}\label{Inst_Igor}{}$^{*_{\theInstitut}}$}

\vskip1cm
\parbox{.9\textwidth}{
\parbox{.4\textwidth}{
\centerline{$^{*_{\ref{Inst_Andre}}}$ Institut f\"ur Physik~~}
\centerline{                           Humboldt Universit\"at}
\centerline{                           D-12489 Berlin, Germany}
\centerline{                           Newtonstr. 15}
\centerline{                           miemiec@physik.hu-berlin.de}                           
}\hfill
\parbox{.4\textwidth}{\centerline{$^{*_{\ref{Inst_Igor}}}$~  Racah Institute of Physics~~~~}
\centerline{                           The Hebrew University}
\centerline{                           Jerusalem 91904}
\centerline{                           Israel}
\centerline{                           igorsc@phys.huji.ac.il}
}
}

\vspace{1.0truecm}
\begin{abstract}
 \noindent
Starting from a nonlinear realisation of eleven dimensional 
supergravity based on the group $G_{11}$, whose generators 
appear as low level generators of $E_{11}$, we present a super 
extended algebra, which leads to a covariant derivative of 
spinors identical to the Killing spinor equation of this theory.
A similar construction leads to the Killing spinor equation of
$N=1$ pure supergravity in ten dimensions.
\end{abstract}
\bigskip \bigskip

\newpage

\section{Introduction}

The search for hidden symmetries of eleven dimensional supergravity 
(M-theory) \cite{Cremmer:1978km}\footnote
{\label{Conventions}
 Our conventions differ from those in \cite{Cremmer:1978km} by 
 rescaling all gauge fields by a factor of $1/2$, working with a 
 mostly plus signature and replacing $\Gamma^a$ by $i\,\Gamma^a$.
}
has a long history \cite{D'Auria:nx}. In the context of 
supergravities' relation to string theories the existence of hidden  
symmetries realising U-duality was first conjectured in 
\cite{Hull:1994ys}. Identifying the (hidden) symmetry group of a 
given supergravity theory is of vast interest for understanding its
properties; it is essential for finding solution generating techniques, 
but also for pinpointing the whole issue of dualities that apparently
relate different supergravities to each other.\\
\noindent 
Several approaches to manifest these extra symmetries have been 
developed in the past. In \cite{Cremmer:1997ct,Cremmer:1998px} it was 
shown that the non-gravitational degrees of freedom of almost all 
supergravity theories can be described as a non-linear realisation 
which made part of the  hidden symmetries manifest. The result was 
obtained by using the doubled field method. The additional degrees of 
freedom introduced by doubling of the fields are projected out by 
the equations of motion of the gauge fields which take the form 
of twisted selfduality conditions. 
The generators of this coset construction were inert under Lorentz
transformations, and as such it is difficult to extend this method
straightforwardly to include gravity or fermions. In \cite{We00} 
it was shown that the entire bosonic sector of eleven- and ten 
dimensional IIA supergravity could be formulated as a non-linear 
realisation. In this way of proceeding, gravity is treated on an 
equal footing with the gauge fields and thus is naturally built in. 
The method of non-linear realisations has consequently been shown to 
extend to other gravity theories 
\cite{Lambert:2001gk,Schnakenburg:2002xx,Schnakenburg:2004vd}. 
However, it did not include the fermionic degrees of freedom. \\
\noindent
The non-linear realisation of M-theory of \cite{We00} is based on the 
nonsimple algebra $G_{11}$ with group element 
\begin{eqnarray}\label{GroupElement}
   g_B &=& e^{x^\mu P_\mu}\,
           e^{\kappa_{a}{}^bK^a{}_b}\,
                      \exp\left(\, 
                          \frac{1}{3!}\,A_{c_1\ldots c_3}R^{c_1\ldots c_3}
                          \,+\,
                          \frac{1}{6!}\,A_{c_1\ldots c_6}R^{c_1\ldots c_6}\,
                      \right)            
\end{eqnarray}
that was shown to generate the covariant structure of the bosonic fields 
(subscript $B$) and their equations of motion. The group $G_{11}$ is 
defined by the algebra (only non-trivial commutators displayed)
\begin{align}
 &
   [\,K^{a}{}_b,\,K^c{}_d\,] ~=~ \delta^c_b\,K^a{}_d\,-\,\delta^a_d\,K^c{}_b 
 & 
 &
   [\,K^a{}_b,\,P_c\,] ~=~ -\,\delta^a_c\,P_b\hspace{1cm}\label{KP=P} 
 &\\[1ex]
 &
   [\,K^a{}_b,\,R^{c_1\ldots c_3}\,] ~=~ 3\,\delta^{[c_1}_b\,R^{c_2c_3]a} 
 &
 &
   [\,K^a{}_b,\,R^{c_1\ldots c_6}\,] ~=~ -\,6\,\delta^{[c_1}_b\,R^{c_2\ldots
                                           c_6]a} &\label{KR=R}\\[1ex]
 & 
   [\,R^{c_1\ldots c_3}\,,R^{c_4\ldots c_6}\,] ~=~ c_{3,3}\,R^{c_1\ldots
  c_6}\label{R3R3=R6} 
 &
 &
 &
\end{align}
and the Cartan form of the coset ($h_{ab}\,=\,\kappa_{(ab)}$, 
$\kappa_{[ab]}\,=\,0$) reads
\begin{eqnarray}\label{BosonicCartanform}
  g_B^{-1}dg_B 
  &=& dx^\mu
      \left\{\vbox{\vspace{2.5ex}}\right.
           P_\mu \,+\, 
           \left(e^{-h}\,\partial_\mu\,e^{h}\right)_a{}^b\,K^a{}_{b} \,+\,
           \frac{1}{4!}\,(4\,\tilde{D}_{[\mu}A_{c_1c_2c_3]})\,R^{c_1c_2c_3} 
           \,+\,\ldots\\
  &&\hspace{38ex}\ldots\,+\,\frac{1}{7!}\,(7\,\tilde{D}_{[\mu}A_{c_1\ldots c_6]})\,R^{c_1\ldots c_6}
      \left.\vbox{\vspace{2.5ex}}\right\}\nonumber   
\end{eqnarray}
with
\begin{eqnarray*}
   \tilde{D}_{\mu}A_{c_1c_2c_3} 
              &=& \partial_{\mu}A_{c_1c_2c_3} \,+\,
                  \left(\,
                          \left(e^{-h}\,\partial_\mu\,e^{h}\right)_{c_1}{}^b 
                           A_{bc_2 c_3}
                    \,+\, \left(e^{-h}\,\partial_\mu\,e^{h}\right)_{c_2}{}^b 
                           A_{c_1 b c_3}
                    \,+\, \left(e^{-h}\,\partial_\mu\,e^{h}\right)_{c_3}{}^b 
                           A_{c_1c_2 b}\,
                  \right)\\
   \tilde{D}_{\mu}A_{c_1\ldots c_6} &=& \partial_{\mu}A_{c_1\ldots c_6} 
                                    ~+~ \left(\,
                                                \left(
                                                   e^{-h}\,\partial_\mu\,e^{h}
                                                \right)_{c_1}{}^b\,
                                                A_{b c_2\ldots c_6}
                                                ~+~  
                                                \left(
                                                   e^{-h}\,\partial_\mu\,e^{h}
                                                \right)_{c_2}{}^b\,
                                                A_{c_1 b c_3\ldots c_6}
                                                ~+~ \ldots\right.\\ 
                                                &&\left.\ldots ~+~ 
                                                \left(
                                                   e^{-h}\,\partial_\mu\,e^{h}
                                                \right)_{c_6}{}^b\,
                                                A_{c_1\ldots c_5 b}\,
                                        \right)
                                    ~-~ 10\,A_{[c_1\ldots c_3}\,
                                        \tilde{D}_{|\mu|}A_{c_4\ldots c_6]}\,
                                        c_{3,3}~.
\end{eqnarray*}
The antisymmetry of the indices indicated in \eqref{BosonicCartanform} 
is not obtained automatically by the procedure outlined so far but it 
is the result of a second step, {\it i.e.} making the Cartan form of 
$G_{11}$ simultaneously covariant with respect to the conformal group 
\cite{Borisov:1974bn,We00}. Identifying the resulting objects with the 
two field strengths $G_{a_1\ldots a_4}$ and $F_{a_1\ldots a_7}$ the
equation of motion of the gauge field of M-theory reads 
\begin{eqnarray}\label{HodgeDuality}
     \ast G_{(4)} &=& F_{(7)}~.
\end{eqnarray}
In this way the bosonic gauge sector of eleven dimensional supergravity
is completely described by the covariant field strengths together with a 
geometric equation of motion. 
The second term in eq.~(\ref{BosonicCartanform}) requires special attention.
From the transformation properties of the Cartan form according to the 
transformation $g_B\mapsto g\cdot g_B\cdot h^{-1}$  and the 
structure of the algebra (\ref{KP=P})-(\ref{R3R3=R6}) it follows that 
the shift term of the Cartan form that appears after performing the
transformation is due to the object in front of the antisymmetric part of 
the generators $K^a{}_b$. The same shift term also 
appears in the transformation of the spin connection. So we may assume 
\begin{eqnarray*}
  \left(e^{-h}\,\partial_\mu\,e^{h}\right)_{[ab]} &=& \omega_{\mu ab} 
   ~+~ \Omega_{\mu [ab]}
\end{eqnarray*} 
with $\Omega_{\mu [ab]}$ transforming as a tensor. We made the antisymmetry
of $\Omega_{\mu ab}$ in the latter two indices explicit since we want 
to extend the definition of $\Omega_{\mu ab}$ to denote the tensors in 
front of the symmetric part of $K^a{}_{b}$, too. Due to the inverse Higgs
effect one can put any Cartan form with a homogenous transformation law to
zero without affecting physics \cite{BorOgi}. This allows one to neglect 
$\Omega_{\mu ab}$ and to find relations between $\omega_{\mu ab}$ and 
$(e^{-h}\partial_{\mu}e^{h})_{ab}$  at the same time (see \cite{We00}).  
In the case of M-theory, non-linear 
realisations led to the proposal of a hidden $E_{11}$ symmetry 
\cite{pwest e11}. $E_{11}$ appeared as the simplest Kac-Moody algebra 
which contained the nonsimple algebra of $G_{11}$ but
without the momentum generator.\\
So far the discussion of the hidden symmetries was limited to the 
bosonic sector of the supergravity only. One open problem of the 
$E_{11}$ conjecture is firstly, how to incorporate fermionic degrees
of freedom and secondly, what restrictions this places on the 
corresponding extension of the algebra $G_{11}$. The best possible 
answer would be to treat the bosons and the fermions (gravitino) on the 
same footing, {\it i.e.} to generate from an extended group 
$\tilde{G}_{11}$ the covariant derivative of the gravitino accompanied 
by the fermionic shifts in the bosonic field strengths and the spin 
connection. Avoiding the construction of a group extension it is 
-of course- possible to introduce the fermionic shifts in the bosonic 
field strengths just by hand \cite{Nurmagambetov:2003my}. 
But then it remains unclear how the extended bosonic symmetries 
couple to the fermionic symmetries, {\it i.e.} what the
extended hidden symmetry group actually looks like.
Alternatively, one can try to generalise the fields of the theory to
superfields in super-space aiming to find the fermionic field equations 
by twisted superdualities. This was performed for a two dimensional 
model in \cite{Paulot:2004fm}.\\

\noindent
Historically, the identification of the Kac-Moody algebra that describes 
the hidden symmetries of the theory was performed by an algebra, which 
did not take the role of the momentum operator as a central charge of the
supersymmetry algebra into account. Later the semi-direct product 
of $E_{11}$ and representations of the momentum generator in eleven 
dimensions were considered. The semi-direct product includes non-trivial 
commutators of the momentum generator with the gauge field generators,
which close in the central charges of the supersymmetry algebra in eleven 
dimensions \cite{We03}. Of course, the momentum generator 
itself appears as a central charge of this algebra. \\
In the following we will rather take the semi-direct product of some
low-level generators of $E_{11}$ ({\it i.e.} $G_{11}$) when split into 
representation of $SL(11)$ with a spinor representation of 
$SO(1,10)$. The Lorentz group can be obtained from the gravity line 
by using the Cartan involution (or the temporal involution 
\cite{Englert}). In this way, we are still free to define the 
anti-commutation relations of this fermionic generator with itself and 
we will naturally choose the supersymmetry algebra of the relevant 
supergravity theory. This algebra contains the momentum generator apart 
from the central charges, and so the momentum generator can effectively 
be added via a semi-direct product.
\noindent
The occurrence of a fermionic generator parametrising the coset of a 
non-linear realisation will result in a covariant expression for a
fermionic parameter, and we suggest to identify this covariant 
expression with the Killing spinor equation of the theory under 
consideration. This fermionic parameter is, however, not part of the
fields of the theory and thus we still keep a purely bosonic background 
but yet include supersymmetry into the ansatz of \eqref{GroupElement}.\\ 
Since the spin representation we multiply is connected with the 
$SO(1,10)$ subgroup generated by the antisymmetric combination of the 
generators $K^{a}{}_b$ of $SL(11)$, we just keep these and throw the 
symmetric combination away (avoiding topological difficulties 
\cite{Keurentjes}). We will partly answer the question as to 
whether there exists an extension of the algebra of $G_{11}$ by a 
fermionic generator $Q_{\dot{\alpha}}$ generating a parameter 
$\varepsilon^{\dot{\alpha}}$, so that the Cartan form finally produces 
a covariant derivative of $\varepsilon^{\dot{\alpha}}$,
\begin{eqnarray}\label{CovDerGravitino}
  \hat{D}_\mu\varepsilon^{\dot{\alpha}} &=& 
       \left(\,\partial_\mu\varepsilon^{\dot{\alpha}}\,-\,\frac{1}{4}\,
                            \omega_{\mu ab}\,
                            \Gamma^{ab}\,\varepsilon^{\dot{\alpha}}\,\right)
                         -\frac{1}{2\cdot 144}
                           \left(
                                 \Gamma^{\alpha\beta\gamma\delta}{}_{\mu}
                                  -8\,\Gamma^{\beta\gamma\delta}
                                     \delta_\mu^\alpha
                           \right)\,\varepsilon^{\dot{\alpha}}\,
                           G_{\alpha\beta\gamma\delta}~, 
\end{eqnarray}
identical to the the Killing spinor equation. Algebraically this problem 
is closely related to the full program of consistently including fermions
into a nonlinear realisation and find them to be Goldstonian. The 
difference is merely that we do not have to consider the fermionic shifts 
induced on the bosonic fields. One should note the conceptual difference 
to \cite{D'Auria:nx}, whose Cartan form contains the potentials 
$A_{c_1\ldots c_3}$ and $A_{c_1\ldots c_6}$ and not the field strengths 
$G_{(4)}$ and $F_{(7)}$.

\section{Supersymmetrisation}

The ansatz for the superalgebra is mainly fixed by the structure of the 
covariant derivative we want to generate. Nevertheless one has to make 
some choices. There are several hints contained in the literature as to 
how the supersymmetric extension might look like 
\cite{D'Auria:nx,We00,vanHolten:1982mx}. For different reasons all 
of these three papers had to take a second, ``unphysical'' spinorial 
generator $\tilde{Q}_{\dot{\alpha}}$ into account. We call this 
generator ``unphysical'' since we do not identify the operator which arises 
in front of $\tilde{Q}_{\dot{\alpha}}$ with the covariant derivative 
of a physical quantity. In fact, in our calculation we 
observed the need for a second fermionic generator, too. We will 
discuss the technical reason below. The most convincing heuristic
argument for the second fermionic generator in the approach via 
nonlinear realisations is derived from the closure of $G_{11}$ with 
the conformal group. The conformal group in $d\,=\,11$ is isomorphic to 
$SO(2,11)$, whose lowest irreducible spin representation is of 
dimension $2^6\,=\,2\cdot 32$, {\it i.e.} twice the amount of the spinor
representation in $d\,=\,11$. This superalgebra was explicitly constructed 
in \cite{vanHolten:1982mx}. \\    
It was laid out in the introduction that the covariant field 
strengths, and in particular their antisymmetrisation, are only found 
after taking the closure with the conformal group. It therefore appears
to be natural to include two fermionic generators $Q_{\dot{\alpha}}$ 
and $\tilde{Q}_{\dot{\alpha}}$ and multiply the group element in 
\eqref{GroupElement} from the right by 
\begin{eqnarray}
   g_\varepsilon &=& e^{
                           \varepsilon^{\dot{\alpha}}\,(\,Q_{\dot{\alpha}}
                           \,+\,\tilde{Q}_{\dot{\alpha}}\,)
                         }~.
\end{eqnarray}
The new Cartan form becomes:
\begin{eqnarray*}
    {\mathcal{A}} &=&  g_\varepsilon^{-1}\,d \,g_\varepsilon
                   ~+~ g_\varepsilon^{-1}\,
                      \left(g_B^{-1}\,d\,g_B\right)\,
                      g_\varepsilon~.
\end{eqnarray*}
We work out the first term on the right hand side using 
eq.~(\ref{d_Expansion}) ending up with an expansion of the form 
\begin{eqnarray*}
     g_\varepsilon^{-1}\,d \,g_\varepsilon &=&  d\varepsilon^{\dot{\alpha}}\,(\,Q_{\dot{\alpha}}
         \,+\,\tilde{Q}_{\dot{\alpha}}\,)
     \,-\, \frac{1}{2}\,[\,\varepsilon^{\dot{\alpha}}\,
                         (\,Q_{\dot{\alpha}}\,+\,\tilde{Q}_{\dot{\alpha}}\,),\,
                        d\varepsilon^{\dot{\beta}}\,
                        (\,Q_{\dot{\beta}}\,+\,\tilde{Q}_{\dot{\beta}}\,)\,]
     \,+\, \ldots 
\end{eqnarray*}
where we have only expanded up to second order since we will soon find
that due to the Jacobi identities commutators with more than two fermionic 
generators vanish (see remark \ref{Rem3Q} on page \pageref{Rem3Q}).
For the 
second bit in the expression  ${\mathcal{A}}$ we introduce a shorthand 
notation. We set  $~g_B^{-1}\,d\,g_B\,=\,\sum_{i=1}^4\,(\ldots)\,{\mathfrak{G}}_i~$
with ${\mathfrak{G}}_i\,\in\,\{\,P_a,\,K^a{}_b,\,R^{c_1c_2c_3},\,R^{c_1\ldots c_6}\}$ where the dots in brackets 
refer to the prefactors in eq.~(\ref{BosonicCartanform}) determined 
in the last paragraph. Then it reads
\begin{eqnarray*}
 g_\varepsilon^{-1}\,
                      \left(g_B^{-1}\,d\,g_B\right)\,
                      g_\varepsilon &=&  
              g_\varepsilon^{-1}\,
           \left(\sum_{i=1}^4\,(\ldots )\,\mathfrak{G}_i\,\right)\,
             g_\varepsilon
\end{eqnarray*}
and using eq.~(\ref{B_Expansion}) we obtain for each of the four individual 
contributions  an expansion of the type
\begin{eqnarray*}
     g_\varepsilon^{-1}
     {\mathfrak{G}}_i\;
     g_\varepsilon
   &=& {\mathfrak{G}}_i 
   ~-~ [\,
           \varepsilon^{\dot{\alpha}}\,(\,Q_{\dot{\alpha}}\,+\,\tilde{Q}_{\dot{\alpha}}\,),\,{\mathfrak{G}}_i\,
       ]
   ~+~ \frac{1}{2!}\,
       [\,
           \varepsilon^{\dot{\beta}}\,(\,Q_{\dot{\beta}}\,+\,\tilde{Q}_{\dot{\beta}}\,),\,
          [\,
              \varepsilon^{\dot{\alpha}}\,(\,Q_{\dot{\alpha}}\,+\,\tilde{Q}_{\dot{\alpha}}\,),\,
              {\mathfrak{G}}_i\,]\,
       ]
   ~+~ \ldots
\end{eqnarray*}

\noindent
It is feasible to organise the expansions by the power in the fermionic 
parameter $\varepsilon^{\dot{\alpha}}$ they contain, {\it i.e.} 
\begin{eqnarray*}
         {\mathcal{A}} &=& \sum\limits_{k=0}^\infty\,{\mathcal{A}}^{(k)}, 
         \quad\quad\quad 
         {\mathcal{A}}^{(0)}\,=\,(g_B^{-1}\,d\,g_B)~.
\end{eqnarray*} 
At zeroth order we just recover the purely bosonic elements of the Cartan
form. Formally, the expansions goes all the way up to infinity depending 
on our choice of (anti)commutation relations of the fermionic generators. 
For the super algebra we are going to use it will terminate at 
$k\,=\,3$.

\subsection{Linearised Analysis, i.e. ${\mathcal{O}}({\varepsilon}^2)$}

\noindent
To first order in $\varepsilon^{\dot{\alpha}}$ the Cartan form looks like 
\begin{eqnarray*}
  {\mathcal{A}}^{(1)}
                  &=& d\varepsilon^{\dot{\alpha}}\,(\,Q_{\dot{\alpha}}\,+\,\tilde{Q}_{\dot{\alpha}}\,)
                  ~-~ \varepsilon^{\dot{\alpha}}\,
                      \sum\limits_{i=1}^{4}\,       
[\,(\,Q_{\dot{\alpha}}\,+\,\tilde{Q}_{\dot{\alpha}}\,),\,(\ldots )\,{\mathfrak{G}}_i\,]
\end{eqnarray*}
Now we evaluate the terms linear in $\varepsilon$ step by step. Because 
of the $\mathbb{Z}_2$-grading of a superalgebra, the commutators between
fermionic and bosonic generators can only yield fermionic generators. They
additionally have to fulfil the super Jacobi identities. We set \footnote
{
   Lorentz generator $J^{ab}\,=\,2\cdot K^{[ab]} \;\;\Rightarrow\;\;
          k^{[ab]}\,=\,\frac{1}{4}\Gamma^{ab}$.
}\label{footnote2} 
\begin{align}
         [\,Q_{\dot{\alpha}},\,K^{[bc]}\,] 
         &~=~ \phantom{-}\;\;
              (k^{[bc]})_{\dot{\alpha}}{}^{\dot{\beta}}\,
              Q_{\dot{\beta}} &
         [\,\tilde{Q}_{\dot{\alpha}},\,K^{[bc]}\,] 
         &~=~ \phantom{-}\;\;
              (k^{[bc]})_{\dot{\alpha}}{}^{\dot{\beta}}\,
              \tilde{Q}_{\dot{\beta}}\nonumber\\{}
         [\,Q_{\dot{\alpha}}{},\,R^{c_1c_2c_3}\,] 
         &~=~ \delta\,
              (\Gamma^{c_1c_2c_3})_{\dot{\alpha}}{}^{\dot{\beta}}\,
              \tilde{Q}_{\dot{\beta}} &
         [\,\tilde{Q}_{\dot{\alpha}}{},\,R^{c_1c_2c_3}\,] 
         &~=~ \kappa\,
              (\Gamma^{c_1c_2c_3})_{\dot{\alpha}}{}^{\dot{\beta}}\,
              Q_{\dot{\beta}}\label{11d superalgebra}\\{}
         [\,Q_{\dot{\alpha}}{},\,R^{c_1\ldots c_6}\,] 
         &~=~ \frac{2\delta\kappa}{c_{3,3}}\,
              (\Gamma^{c_1\ldots c_6})_{\dot{\alpha}}{}^{\dot{\beta}}\,
              Q_{\dot{\beta}} &
         [\,\tilde{Q}_{\dot{\alpha}}{},\,R^{c_1\ldots c_6}\,] 
         &~=~ \frac{2\delta\kappa}{c_{3,3}}\,
              (\Gamma^{c_1\ldots c_6})_{\dot{\alpha}}{}^{\dot{\beta}}\,
              \tilde{Q}_{\dot{\beta}}\nonumber
\end{align}
where $\delta$ and $\kappa$ are free parameters. Appendix \ref{Appendix1} 
shows that this choice is consistent with the super Jacobis. We note that 
it is essential to observe that in the second line the commutator with 
$R^{a_1a_2a_3}$ exchanges the two different $Q_{\alpha}$ generators!
However, since one of them is non-physical, 
we only display terms proportional to $Q_{\dot{\alpha}}$ which look
\begin{eqnarray}\label{CF1F}
  {\mathcal{A}}^{(1)}
                  &=& dx^\mu\,
                      \left\{\vbox{\vspace{2.5ex}}\right.
                             \partial_{\mu}\varepsilon^{\dot{\alpha}}\,
                             Q_{\dot{\alpha}}
                         ~-~ \varepsilon^{\dot{\alpha}}\,e^b_\mu
                             \underbrace{[\,Q_{\dot{\alpha}},\,P_b\,]}_{0~!}
                         ~-~ \varepsilon^{\dot{\alpha}}\,
                             \omega_{\mu bc}\,
                             [\,Q_{\dot{\alpha}},\,K^{[bc]}\,]\\
                        &&-~ \varepsilon^{\dot{\alpha}}\,
                             \frac{1}{4!}\,(4\,\tilde{D}_{\mu}A_{c_1c_2c_3})\,
                             [\,\tilde{Q}_{\dot{\alpha}},\,R^{c_1c_2c_3}\,]
                         ~-~ \varepsilon^{\dot{\alpha}}\,
                             \frac{1}{7!}\,
                             (7\,\tilde{D}_{\mu}A_{c_1\ldots c_6})\,
                             [\,Q_{\dot{\alpha}},\,R^{c_1\ldots c_6}\,]
                      \left.\vbox{\vspace{2.5ex}}\right\}\nonumber
\end{eqnarray}
and the commutator $[Q,P]$ vanishes due to the Jacobi identity 
Nr. (\ref{Jid_34}) in Appendix \ref{Appendix1}. The further 
simplifications of eq.~(\ref{CF1F}) are straightforward. The only trick 
one has to keep in mind is connected with rewriting the term containing 
the generator $R^{c_1\ldots c_6}$. We have to use the equations of 
motion, {\it i.e.} 
the condition that the two gauge field strengths $G_{(4)}$ and 
$F_{(7)}$ are related by Hodge duality eq.~(\ref{HodgeDuality}), to 
draw the following conclusion:\footnote
{
     $\Gamma_{a_1\ldots
       a_j}\,=\frac{(-1)^\frac{(11-j)(11-j-1)}{2}}{(11-j)!}\,\Gamma_{a_1\ldots a_{11}}\,\Gamma^{a_{j+1}\ldots a_{11}}$ and
     $\Gamma^{0\ldots 10} = {\rm sgn}\{0\ldots 10\}$
}:
%
\begin{eqnarray}\label{HilfsTrick}
   \frac{1}{6!}\,F_{\mu\mu_2\ldots \mu_7}\,
   \Gamma^{\mu_2\ldots\mu_7}
   &=&
   \frac{1}{4!}\,
   \Gamma_{\mu}{}^{\beta_1\ldots\beta_4}\,G_{\beta_1\ldots \beta_4}
\end{eqnarray}
Using this identity we may rewrite the contribution delivered by the 6-form 
potential into:
\begin{eqnarray*}
  \varepsilon^{\dot{\alpha}}\,\frac{1}{7!}\,
  (7\,\tilde{D}_{\mu}A_{c_1\ldots c_6})\,
  [\,Q_{\dot{\alpha}},\,R^{c_1\ldots c_6}\,]
  &=& \frac{2\delta\kappa}{c_{3,3}}\,\varepsilon^{\dot{\alpha}}\,
      \left\{\,
                \frac{1}{7\cdot 4!}\,\Gamma_{\mu}{}^{\beta_1\ldots\beta_4}\,G_{\beta_1\ldots\beta_4}\,
      \right\}\,
      Q_{\dot{\beta}}
\end{eqnarray*}
Inserting this into eq.~\eqref{CF1F} one obtains:
\begin{eqnarray}\label{KSPVersion1}
  {\mathcal{A}}^{(1)}
                &=& dx^\mu\,
                      \left\{\vbox{\vspace{2.5ex}}\right.\,
                         \partial_\mu\varepsilon^{\dot{\alpha}}
                         ~-~ \omega_{\mu bc}\,
                             (k^{[bc]})_{\dot{\beta}}{}^{\dot{\alpha}}\,
                             \varepsilon^{\dot{\beta}}\\
                       &&\hspace{0.8cm}-~ \frac{1}{4!}\,
                             \left(\vbox{\vspace{2.5ex}}\right.\,
                               \left[\,
                                      \frac{2\delta\kappa}{7\,c_{3,3}}\,
                               \right]\,
                               (\Gamma^{c_0\ldots c_3}
                               {}_{\mu})_{\dot{\beta}}{}^{\dot{\alpha}}
                               \,+\,\left[\,\kappa\,\right]\,
                               \delta_{\mu}^{c_0}\,
                             (\Gamma^{c_1c_2c_3})_{\dot{\beta}}
                                                {}^{\dot{\alpha}}\,
                             \left.\vbox{\vspace{2.5ex}}\right)\,
                             G_{c_0c_1c_2c_3}\,
                             \varepsilon^{\dot{\beta}}\,
                      \left.\vbox{\vspace{2.5ex}}\right\}\,
                      Q_{\dot{\alpha}}\nonumber
\end{eqnarray}
Up to this point we have not fixed any of the free parameters 
appearing in our predictions for the equations of motion of
11d supergravity. Now we want to fix the three free parameters
$c_{3,3}$, $\delta$ and $\kappa$ in a way, which finally produce 
the correct equation of motion for the gauge field strength 
$G_{(4)}$ and the Killing spinor equation. We have to choose
\begin{align}\label{Parameters}
   \hspace{3cm}  c_{3,3} &~=~ 1, & \frac{2\delta\kappa}{7\,c_{3,3}} &~=~
   \frac{1}{12}, & \kappa &~=~ -\,\frac{8}{12}~. \hspace{3cm}
\end{align} 
These constraints lead to  $\delta\,=\,-\,\frac{7}{16}$.
The complete Cartan form for the fermionic generator becomes 
\begin{eqnarray*}
       {\mathcal{A}}^{(1)}
       &=& dx^\mu\,(\hat{D}_\mu^{(0)}\varepsilon)^{\dot{\alpha}}
           Q_{\dot{\alpha}}
       ~+~ dx^\mu\,(\hat{\Delta}_\mu^{(0)}\varepsilon)^{\dot{\alpha}}
           \tilde{Q}_{\dot{\alpha}}
\end{eqnarray*}
with $(D_\mu^{(0)}\varepsilon)^{\dot{\alpha}}$ the Killing spinor equation of
eq.~(\ref{CovDerGravitino}) and 
$(\Delta_\mu^{(0)}\varepsilon)^{\dot{\alpha}}$ the operator in front of 
the "unphysical" generator $\tilde{Q}_{\dot{\alpha}}$ which is of no 
importance for the physical quantities.\\ 

\noindent
The higher order corrections can be computed similarly. At next order 
${\mathcal{O}}(\varepsilon^3)$ we obtain 
\begin{eqnarray}
    {\mathcal{A}}^{(2)} 
     &=& -\,\frac{1}{2}\,\varepsilon^{\dot{\alpha}}\,
         (\hat{D}^{(0)}_\mu\varepsilon)^{\dot{\beta}}\,
         \{\,Q_{\dot{\alpha}},\,Q_{\dot{\beta}}\,\}\,dx^\mu
     ~-~ \frac{1}{2}\,\varepsilon^{\dot{\alpha}}\,
         (\hat{\Delta}^{(0)}_\mu\varepsilon)^{\dot{\beta}}\,
         \{\,\tilde{Q}_{\dot{\alpha}},\,\tilde{Q}_{\dot{\beta}}\,\}\,dx^\mu
         \nonumber\\
     && -\,\frac{1}{2}\,\varepsilon^{\dot{\alpha}}\,
         \left(\,
                   (\hat{D}^{(0)}_\mu\varepsilon)^{\dot{\beta}}\,
                   \{\,\tilde{Q}_{\dot{\alpha}},\,Q_{\dot{\beta}}\,\}
                   \,+\,
                   (\hat{\Delta}^{(0)}_\mu\varepsilon)^{\dot{\beta}}\,
                   \{\,Q_{\dot{\alpha}},\,\tilde{Q}_{\dot{\beta}}\,\}
         \right)\,dx^\mu~.
\end{eqnarray}
which is an expression in the various central charges 
(~cf.~(\ref{ZentrLadung})~). All terms ${\mathcal{A}}^{(k>2)}$ vanish 
due to the Jacobi identities (A.\ref{Jid_34})-(A.\ref{Jid_37}) and remark
\ref{Rem3Q} on page \pageref{Rem3Q}.
Putting all the results for ${\mathcal{A}}$ together one obtains 
\begin{eqnarray}
   {\mathcal{A}} ~=~ \sum\limits_{i=1}^\infty {\mathcal{A}}^{(i)}
                 &=& (g_B^{-1}dg_B)
                 ~+~ dx^\mu\,(\hat{D}^{(0)}_\mu\varepsilon)^{\dot{\alpha}}\,
                     \left(\,
                             Q_{\dot{\alpha}}~-~\frac{1}{2}\,
                             \varepsilon^{\dot{\beta}}\,
                             \{\,Q_{\dot{\alpha}},\,
                             Q_{\dot{\beta}}\,\}\,
                     \right)\nonumber\\
                 &&\hspace{9ex}~
                 +~ dx^\mu\,
                    (\hat{\Delta}^{(0)}_\mu\varepsilon)^{\dot{\alpha}}\,
                    \left(\,
                              \tilde{Q}_{\dot{\alpha}}~-~
                              \frac{1}{2}\,\varepsilon^{\dot{\beta}}\,
                              \{\,\tilde{Q}_{\dot{\alpha}},\,
                              \tilde{Q}_{\dot{\beta}}\,\}\,
                    \right)\\
&& \hspace{9ex}-\,\frac{1}{2}\,\varepsilon^{\dot{\alpha}}\,
         \left(\,
                   (\hat{D}^{(0)}_\mu\varepsilon)^{\dot{\beta}}\,
                   \{\,\tilde{Q}_{\dot{\alpha}},\,Q_{\dot{\beta}}\,\}
                   \,+\,
                   (\hat{\Delta}^{(0)}_\mu\varepsilon)^{\dot{\beta}}\,
                   \{\,Q_{\dot{\alpha}},\,\tilde{Q}_{\dot{\beta}}\,\}
         \right)\,dx^\mu~.\nonumber
\end{eqnarray}
It is important to notice that there is a correction term to the bosonic  
vielbein in front of the momentum generator coming from the anticommutator 
\{Q,\,Q\}, which is proportional to $\hat{D}^{(0)}_\mu\varepsilon$. If one 
imposes the Killing spinor equation the bosonic vielbein is left unchanged. 
This is an a 
posteriori argument for the identification of the Killing spinor equation 
and our notion of ``physical'' and ``unphysical'' fermionic generators.

\section{$N=1$ pure supergravity}

A similar construction as the one before can also be used to find
the Killing spinor equation of $N=1$ pure supergravity in
ten dimensions \cite{Chamseddine:1980}\footnote
{
 The differences in the conventions to \cite{Chamseddine:1980} 
 are the same as described in footnote \ref{Conventions}.
}.
The group that was used to construct the covariant objects of the bosonic 
sector of  the theory was spelled out in \cite{Schnakenburg:2004vd}, and 
was called $G_{I}$. The group element is taken to be
\begin{equation}
   g=e^{x^\mu\,P_\mu}\,e^{h_a{}^b\,K^a{}_b}\,
     e^{\frac{1}{8!}\,A_{a_1\ldots a_8}\,R^{a_1\ldots a_8}}\,
     e^{\frac{1}{6!}\,A_{a_1\ldots a_6}\,R^{a_1\ldots a_6}}\,
     e^{\frac{1}{2!}\,A_{a_1a_2}\,R^{a_1a_2}}\,e^{A\,R}
\end{equation}
and the commutators of the generators satisfy relations analogous to 
eq.~(\ref{KP=P})-(\ref{KR=R}) but with a new set of gauge field 
commutators, whose algebra is given by
\begin{equation}\label{TypeIGaugeAlgebra}
   [\,R,\, R^{a_1\ldots a_p}\,] ~=~ c_p\,R^{a_1\ldots a_p},\quad 
   [\,R^{a_1a_2},\, R^{a_3\ldots a_8}\,] ~=~ c_{2,6}\,R^{a_1\ldots a_8},\qquad 
   c_2\,=\,-\,c_6 \,=\,c_{2,6} \,=\,\frac{1}{2}~.
\end{equation}
The corresponding field strengths (closure with the conformal group yields 
antisymmetric tensors as described before) are 
\begin{eqnarray}
       F_{a_1} &=& \partial_{a_1}\,A
                \label{F_1} \\
    F_{a_1a_2a_3} &=& e^{-\frac{A}{2}}\,
                         (3\,\partial_{[a_1}\,A_{a_2a_3]})
                      \label{F_3}\\
    F_{a_1\ldots a_7} &=& e^{\frac{A}{2}}\,(7\,
                          \partial_{[a_1}A_{a_2\ldots a_7]})
                          \label{F_7}\\
    F_{a_1\ldots a_9} &=& 9\,
                          \left(\,
                                   \partial_{[a_1}A_{a_2\ldots a_9]}
                                   \,-\,
                                   7\cdot 2\,A_{[a_1a_2}\partial_{a_3}
                                   A_{a_4\ldots a_9]}\,
                          \right)
                          \label{F_9}
\end{eqnarray}
with the two first order equations of motion
\begin{align}\label{EOMTypeI}
   \hspace{4cm}
   \ast F^{(3)} &= F^{(7)}~, &  
   \ast F^{(1)} &= F^{(9)}~.
   \hspace{4cm}
\end{align} 
In our notation the two Killing spinor equations read 
\begin{eqnarray}
   \delta\psi_\mu &=& D_\mu\varepsilon ~+~ \frac{1}{72}\,
                      \left(\,
                                \Gamma^{\nu\rho\sigma}{}_\mu
                                \,-\,9\delta_\mu^\nu\,\Gamma^{\rho\sigma}\,
                      \right)\,\varepsilon\,F_{\nu\rho\sigma}\label{KSPTypeI}\\
   \delta\chi     &=& \sqrt{\frac{1}{8}}\,
                      \left(\,
                              F_\mu\,\Gamma^\mu\,\varepsilon ~-~
                              \frac{1}{12}\,
                              \Gamma^{\nu\rho\sigma}\,
                              F_{\nu\rho\sigma}\,\varepsilon\,
                      \right)\label{AlgKSPTypeI}
\end{eqnarray}
As in the case of eleven dimensional supergravity treated previously we 
enhance the group element by a fermionic generator 
$\exp(\varepsilon^\alpha\, Q_\alpha)$ from the right. The extension of 
the algebra is defined by commutation relations similarly 
to \eqref{11d superalgebra} and reads
\begin{equation}
\begin{split}
   [Q_{\dot\alpha},\, R^{a_1a_2}] & = s_2 
    (\Gamma^{a_1a_2})_{\dot\alpha}{}^{\dot\beta} Q_{\dot\beta}\\
   [Q_{\dot\alpha},\, R^{a_1\ldots a_6}] & = s_6
    (\Gamma^{a_1\ldots a_6})_{\dot\alpha}{}^{\dot\beta}Q_{\dot\beta}
   + q_6
    (\Gamma^{a_1\ldots a_6})_{\dot\alpha}{}^{\dot\beta}\tilde
    Q_{\dot\beta}\label{TypeI superalgebra}\\
   [Q_{\dot\alpha},\, R^{a_1\ldots a_8}] & = q_8
    (\Gamma^{a_1\ldots a_8})_{\dot\alpha}{}^{\dot\beta}\tilde Q_{\dot\beta}~.
\end{split}
\end{equation}
Since previous experience has taught us to introduce a second fermionic 
generator we do it here again and discuss the reason later.\\

\noindent
In eq.~(\ref{TypeI superalgebra}) we have not written down commutators 
of the dilaton generator with the supercharges. Actually, by checking 
the Jacobi identities it is found that the super extension is 
inconsistent with the interpretation of the dilaton generator as an 
element of the Cartan subalgebra.
We are not surprised. The origin of this problem is connected to the 
difficulties with the symmetric part of $K^a{}_b$. In the footnote on page
\pageref{footnote2} we have used the antisymmetric $\Gamma$-matrices to 
parameterise the antisymmetric part of the $K^{ab}$ generators. The
dilaton generator $R$ can be understood from an eleven dimensional 
perspective as a generator built from the trace parts of the eleven 
dimensional $K^a{}_b$. Since we have also realised the gauge generators 
inside the Clifford algebra there is no algebraic possibility to 
realise the $R$ generator inside the Clifford algebra at the same time.
Perhaps there is another mathematical technique to get rid of the dilaton 
but it is unknown to us. Since we do not need the contributions
from the dilaton generator $R$, we have dropped it by hand; we 
have to keep in mind though that finding a closing algebra including 
$R$ needs further consideration.\\

\noindent
As usually, we are mainly interested in those elements that close in 
the untilded $Q_\alpha$. 
The various constants $s_i$ and $q_i$ are not linearly independent but 
have to be chosen such that the Jacobi identities close. We have to 
formally define also commutators 
\begin{equation}
\begin{split}
   [\tilde Q_{\dot\alpha},\, R^{a_1a_2}] & = \tilde q_2 
    (\Gamma^{a_1a_2})_{\dot\alpha}{}^{\dot\beta} \tilde Q_{\dot\beta}\\
   [\tilde Q_{\dot\alpha},\, R^{a_1\ldots a_6}] & = \tilde s_6
    (\Gamma^{a_1\ldots a_6})_{\dot\alpha}{}^{\dot\beta}Q_{\dot\beta}
   + \tilde q_6
    (\Gamma^{a_1\ldots a_6})_{\dot\alpha}{}^{\dot\beta}\tilde Q_{\dot\beta}\\
   [\tilde Q_{\dot\alpha},\, R^{a_1\ldots a_8}] & = \tilde s_8
    (\Gamma^{a_1\ldots a_8})_{\dot\alpha}{}^{\dot\beta}\,Q_{\dot\beta}~.
\end{split}
\end{equation}
The Jacobi identities put the constraint below on the free coefficients  
above:
\begin{eqnarray}
    [\,Q_\alpha,\, [\,R^p,\, R^q\,]\,] & = [\,[\,Q_\alpha,\,R^p\,],\, R^q\,] 
     ~+~ [\,R^p,\,[\,Q_\alpha,\, R^q\,]\,]
\end{eqnarray}
and the one with $Q$ and $\tilde Q$ exchanged. We note that in this case the 
commutation relations are actually
carried by the gamma-matrices if we totally antisymmetrise the indices. 
Evaluated  on the totally antisymmetric part of the Clifford algebra it 
gives the following constraints:
\begin{center}
\refstepcounter{table}
\label{TypeIConstraints}
\begin{tabular}{|c|c|}
\hline
       & \\[-1ex]
 (p,q) & $[\,Q_\alpha,\, [\,R^p,\, R^q\,]\,]$ - constraint\\[0.5ex]
\hline
       &      \\[-1ex]
 (2,2) & none \\
 (2,6) & $c_{2,6}\,q_8=q_6\,(\,s_2-\tilde q_2\,)$\\
 (2,8) & $0\,=\,q_8\,(s_2-\tilde q_2)$\\
 (6,6) & none\\
 (6,8) & $0\,=\,q_8\,(s_6-\tilde q_6)$ \& $0\,=\,q_6\tilde s_6\,(s_2-\tilde q_2)$\\
 (8,8) & none\\[1ex]
\hline
\end{tabular}
\end{center}
This must be solved by setting
\begin{eqnarray}
    s_2 ~=~ \tilde q_2~, \qquad q_8~=~0
\end{eqnarray}
The constraints from the  $[\,\tilde Q_\alpha,\, [\,R^p,\, R^q\,]\,]$ 
Jacobi identity are completely analogous. 
The remaining two constants $s_2,\, s_6$ are unconstrained from
this point of view and can be chosen as to generate the Killing spinor 
equation of pure supergravity in ten dimensions.
The Cartan form becomes
\begin{eqnarray}
   g^{-1}\,d\,g &=& g_B^{-1}dg_B ~+~ dx^a 
   \,\left(\,
            \partial_a\varepsilon^\beta \,-\, 
            \frac{s_2}{3!}\,F_{aa_2a_3}\,(\Gamma^{a_2a_3})_\alpha{}^\beta\,
            \varepsilon^{\alpha}\,-\,
            \frac{s_6}{7!}\,F_{aa_2\cdots a_7}\,
            (\Gamma^{a_2\cdots a_7})_\alpha{}^\beta\,\varepsilon^{\alpha}\,
   \right)\,Q_\beta\nonumber  \\
   && +\, dx^a\, 
   \left(\, 
            -\,\frac{q_6}{7!}\,F_{aa_2\cdots a_7}\,
            (\Gamma^{a_2\cdots a_7})_\alpha{}^\beta\,\varepsilon^{\alpha} 
            \,-\,
            \frac{q_8}{9!}\,F_{aa_2\cdots a_9}\,
            (\Gamma^{a_2\cdots a_9})_\alpha{}^\beta\,\varepsilon^{\alpha}\, 
   \right)\,\tilde Q_\beta + \ldots~.\label{Cartan form N=1}
\end{eqnarray}
Using the equations of motion of eq.~(\ref{EOMTypeI}) as in the last 
section to get rid of the dual field strengths\footnote
{
   $\Gamma_{a_1\ldots a_j} \,=\,\frac{1}{(10-j)!}\,
                              \Gamma_{a_1\ldots
   a_{j}b_1\ldots b_{10-j}}\,\Gamma^{b_{10-j}\ldots b_{1}}$

},
\begin{eqnarray*}
   \frac{1}{p!}\,F_{c_0c_1\ldots c_p}\,\Gamma^{c_1\ldots c_p}
   &=& \frac{(-1)^{p}\,(-1)^{\frac{(10-p)(9-p)}{2}}}{(10-p-1)!}\,
       \Gamma_{11}\,G^{a_{p+1}\ldots a_9}\,\Gamma_{c_0a_{p+1}\ldots a_{9}}~,
\end{eqnarray*}
the expression in front of $Q_{\beta}$ in \eqref{Cartan form N=1} 
simplifies to 
\begin{eqnarray*}
   &&      \partial_a\varepsilon^\beta\,+\,\frac{1}{72}\,
           \left(\,
                       \frac{12\,s_6}{7}\,
                          (\Gamma^{a_1a_2a_3}{}_{\mu})_\alpha{}^\beta
                     \,-\,12\,s_2\,\delta_a^{a_1}
                           (\Gamma^{a_2a_3})_\alpha{}^\beta\,
           \right)\,\varepsilon^{\alpha}\,F_{a_1a_2a_3}
\end{eqnarray*}
Comparison with the Killing spinor equation eq.~(\ref{KSPTypeI}) fixes 
\begin{eqnarray}
            s_6 &=& \frac{7}{12},\quad s_2 ~=~ \frac{3}{4}~.
\end{eqnarray}
Finally, let us take a look at the expression that builds up
in front of $\tilde Q$ in 
eq.~(\ref{Cartan form N=1}). After using the equations of motion 
eq.~(\ref{EOMTypeI}) we found indications  that this could be 
connected with the algebraic Killing spinor equation 
eq.~(\ref{AlgKSPTypeI}). If one contracts this expression with 
$\Gamma^{a}$ one obtains
\begin{equation}
\begin{split}
    {\rm in}\,\, \tilde Q\,: & \quad
   -\,\frac{q_6}{3!}(\Gamma^{b_1b_2b_3})_\alpha{}^\beta\,\varepsilon^{\alpha}\,  
   F_{b_1b_2b_3} \,+\,q_8\,F_b
   (\Gamma^b)_\alpha{}^\beta\,\varepsilon^{\alpha}
\end{split}
\end{equation}
which is exactly the expected structure. 
Due to the gauge algebra of $G_I$ in (\ref{TypeIGaugeAlgebra}) which requires 
$c_{2,8}\equiv 0$ 
we have to place the stopper $q_8=0$ (see table) which deletes the 
second term of the above equation. In $D_8^{+++}$, however, 
$c_{2,8}\equiv 0$ is not required anymore, and thus $q_8 \neq 0$ is
possible \cite{Axel}. Obviously the solution to the dilaton puzzle holds the key
to the completion of the picture.\\ 
The relation of this super extension to the one defined in the case of 
eleven dimensional supergravity is not understood.

\section{Conclusions}

We have shown that part of the original $G_{11}$ group used in 
\cite{We00} to define M-theory as a nonlinear realisation possesses 
an extension whose Cartan form produces a covariant derivative of 
spinors identical to the Killing spinor equation of eleven dimensional 
supergravity. It is appealing that the structure of the Killing spinor 
equation is inevitably generated by the group structure. On the other
hand this method is not yet expected to give the full super covariant
objects like the supercovariant field strengths including the fermionic 
shifts.\\
A similar construction was used for $N=1$ pure supergravity but runs 
into difficulties because of the dilaton generator that arises by 
dimensional reduction from eleven dimensions on a torus.\\
It would be interesting to see how dimensional reduction can be made
consistent with a super algebra, since the problems with the dilaton
generator are generic. We expect that the approach presented here can
be straightforwardly applied to other supergravity theories and 
other low-level expansions of very-extended Lie algebras (see 
\cite{Axel}) and might hold some clues about the above stated problem.
In this case it could be helpful to classify all maximal supersymmetric
version of supergravities \cite{Farrill,Andre} but also other 
solutions that preserve different amounts of supersymmetry.\\
A crucial point turned out to be the need to introduce at least two 
fermionic generators. Working with just one generator it is not 
possible to fix the free parameters of eq.~(\ref{Parameters}) 
consistently. The doubling of the fermionic generators is interesting 
of itself and related to the identification of positive and negative
roots of a super algebra.\\
Finally, it would be useful to see the relation to other approaches 
assuming  infinite dimensional Kac-Moody algebras as symmetry algebras
of (super)gravity theories \cite{Hermann,Ganor}. Our 
approach of taking the semi-direct product with a spinor representation 
should be applicable to these models, too.

\section{Acknowledgements}
A.M. and I.S. would like to thank P. Fr{\'e} and P. West for discussion at
an earlier stage of the work. A.M. thanks I.~Kirsch for fruitful
discussions. Furthermore would I.S. like to thank B. Julia,
A. Keurentjes, A. Kleinschmidt, and H. Nicolai for interesting 
discussions, explanations and generous hospitality. The work of A.M. is 
supported by the Deutsche Forschungs\-ge\-meinschaft (DFG). The work of 
I.S. is partly supported by BSF - American-Israel Bi-National Science 
Foundation, the Israel Academy of Sciences - Centers of Excellence 
Program, the German-Israel Bi-National Science Foundation, and the 
European RTN-network HPRN-CT-2000-00122.

\begin{appendix}

\section{Jacobi Identities of M-Theory}
\label{Appendix1}
\newcounter{nummer}

In this appendix we list for completeness the set of Jacobi identities 
in the case of eleven dimensional supergravity. The Lie-bracket must 
be understood according to the parity ($\,|\ldots |\,$) of the generators 
as commutators and anticommutators, respectively, {\it i.e.}
\begin{eqnarray*}
   [\,X,\,Y\,] &=& -\,(-1)^{|X|\cdot |Y|}\cdot [\,Y,\,X\,]~.
\end{eqnarray*}
The corresponding Jacobi identity reads
\begin{eqnarray}\label{JacobiIdentity}
  [\,X,\,[\,Y,\,Z\,]\,] 
  &=& [\,[\,X,\,Y\,],\,Z\,]
  ~+~ (-1)^{|X|\cdot |Y|}\,[\,Y,\,[\,X,\,Z\,]\,]~.  
\end{eqnarray}
The consistency of the purely bosonic generators was established in 
\cite{We00}. So we concentrate on the Jacobis containing one or more 
fermionic generators.  In the next section we consider Jacobis containing 
at most one fermionic generator. We have chosen $Q_{\dot{\alpha}}$ but 
the case of $\tilde{Q}_{\dot{\alpha}}$ is totally symmetric.

\subsection{Jacobis with one fermionic generator $Q_{\dot{\alpha}}$}

For commutators of two $R^{c_1\ldots c_3}$ given in eq.~(\ref{R3R3=R6}) 
we make explicit the implicit requirement on the symmetry by introducing 
the anti symmetrisation symbol on the right hand side:
\begin{eqnarray}
   [\,R^{[c_1\ldots c_3},\,R^{d_1\ldots d_3]}\,] 
   &=& c_{3,3}\,R^{[c_1\ldots c_3d_1\ldots d_3]} ~.
\end{eqnarray}
The additional projection onto the totally antisymmetric part
makes the fermionic extension possible, {\it i.e.} the above 
algebra can be realised as the total antisymmetric part in the 
Clifford multiplication.\\ 
In contrast to eq.~(\ref{11d superalgebra}) we define for the 
purpose of shortness
\begin{eqnarray}
 [\,Q_{\dot{\alpha}},\,K^c{}_d\,] 
  &=&  (k^c{}_d)_{\dot{\alpha}}{}^{\dot{\beta}}\,
      Q_{\dot{\beta}}\label{QK}\\{}
 [\,Q_{\dot{\alpha}},\,R^{c_1c_2c_3}\,] 
  &=& \delta\,
       (\Gamma^{c_1c_2c_3})_{\dot{\alpha}}{}^{\dot{\beta}}\,
       Q_{\dot{\beta}}\label{QR3}~.
\end{eqnarray}
The reader may immediately notice that we seem to work with the symmetric 
part of $K^a{}_b$, too. This is correct up to the fact that we don't know 
an explicit realisation of these  $k^c{}_d$ (if it exists at all 
\cite{Keurentjes}). To resolve any ambiguities 
one can restrict the corresponding equation onto the antisymmetric part 
of  $k^c{}_d$, which possess an explicit realization as the generators 
of $\mathfrak{spin}(1,10)$. In Tab.~\ref{OneFermionJacobianIdentities_Q} 
we list all Jacobi identities containing at most one fermionic generator 
by displaying the left hand side of eq.~(\ref{JacobiIdentity}).

\begin{center}
\refstepcounter{table}
\label{OneFermionJacobianIdentities_Q}
\begin{tabular}{|c|l|c|}
\hline
    &      &\\[-1ex]
 Nr.& l.h.s. of eq.~(\ref{JacobiIdentity}) & satisfied ? \\[0.5ex]
\hline
    &      &\\[-1ex]
    \refstepcounter{nummer}\label{Jid_21}(\thenummer) & 
    $[\,Q_{\dot{\alpha}},\,[\,P_b,\,P_c\,]\,]$ &  {\sl trivial}\\
    \refstepcounter{nummer}\label{Jid_22}(\thenummer) & 
    $[\,Q_{\dot{\alpha}},\,[\,K^b{}_c,\,K^d{}_e\,]\,]$ & {\sl cf. proof} \\
    \refstepcounter{nummer}\label{Jid_23}(\thenummer) & 
    $[\,Q_{\dot{\alpha}},\,[\,R^{b_1b_2b_3},\,R^{c_1c_2c_3}\,]\,]$ & 
    {\sl cf. proof} \\
    \refstepcounter{nummer}\label{Jid_24}(\thenummer) & 
    $[\,Q_{\dot{\alpha}},\,[\,R^{b_1\ldots b_6},\,R^{c_1\ldots c_6}\,]\,]$ 
    & {\sl cf. proof}  \\[1ex]
\hline
    &       &\\[-1ex]
    \refstepcounter{nummer}\label{Jid_25}(\thenummer) & 
    $[\,Q_{\dot{\alpha}},\,[\,P_b,\,K^c{}_d\,]\,]$ & {\sl trivial} \\
    \refstepcounter{nummer}\label{Jid_26}(\thenummer) & 
    $[\,Q_{\dot{\alpha}},\,[\,P_b,\,R^{c_1c_2c_3}\,]\,]$ & {\sl trivial} \\
    \refstepcounter{nummer}\label{Jid_27}(\thenummer) & 
    $[\,Q_{\dot{\alpha}},\,[\,P_b,\,R^{c_1\ldots c_6}\,]\,]$ &  {\sl trivial}\\
    \refstepcounter{nummer}\label{Jid_28}(\thenummer) & 
    $[\,Q_{\dot{\alpha}},\,[\,K^b{}_c,\,R^{c_1c_2c_3}\,]\,]$ &  
    {\sl cf. proof}\\
    \refstepcounter{nummer}\label{Jid_29}(\thenummer) & 
    $[\,Q_{\dot{\alpha}},\,[\,K^b{}_c,\,R^{c_1\ldots c_6}\,]\,]$ & 
    {\sl cf. proof} \\
    \refstepcounter{nummer}\label{Jid_30}(\thenummer) & 
    $[\,Q_{\dot{\alpha}},\,[\,R^{b_1b_2b_3},\,R^{c_1\ldots c_6}\,]\,]$ 
    & {\sl cf. proof} \\[1ex]
\hline
\end{tabular}
\center{{\bf Tab.~\thetable~} One fermionic $Q_{\dot{\alpha}}$-generator}
\end{center}

\begin{proof} of Nr.~(\ref{Jid_22})\\

\noindent
\begin{eqnarray*}
   [\,Q_{\dot{\alpha}},\,[\,K^b{}_c,\,K^d{}_e\,]\,] 
   &=& [\,Q_{\dot{\alpha}},\,
          \delta^d_c\,K^b{}_e\,-\,\delta^b_e\,K^d{}_c\,]\,]\\
   &=&   \left\{
                     \delta^d_c\,(k^b{}_e)_{\dot{\alpha}}{}^{\dot{\beta}}
                     \,-\,
                     \delta^b_e\,(k^d{}_c)_{\dot{\alpha}}{}^{\dot{\beta}}\,
          \right\}\,Q_{\dot{\beta}}\\[2ex]{}
   [\,Q_{\dot{\alpha}},\,[\,K^b{}_c,\,K^d{}_e\,]\,] 
   &=& [\,[\,Q_{\dot{\alpha}},\,K^b{}_c\,],\,K^d{}_e\,]\,] ~+~
       [\,K^b{}_c,\,[\,Q_{\dot{\alpha}},\,K^d{}_e\,]\,]\\{}
   &=& ([\,k^b{}_c,\,k^d{}_e\,])_{\dot{\alpha}}{}^{\dot{\gamma}}\,
           Q_{\dot{\gamma}}
\end{eqnarray*}
i.e. 
\begin{eqnarray*}
    ([\,k^b{}_c,\,k^d{}_e\,])_{\dot{\alpha}}{}^{\dot{\beta}}
    &=& \delta^d_c\,(k^b{}_e)_{\dot{\alpha}}{}^{\dot{\beta}}
                     \,-\,
        \delta^b_e\,(k^d{}_c)_{\dot{\alpha}}{}^{\dot{\beta}}
\end{eqnarray*}
forms a representation of $K^a{}_b$.
\end{proof}

\begin{proof} of Nr.~(\ref{Jid_23})\\

\noindent
Using eq.~(\ref{R3R3=R6}) we get
\begin{eqnarray*}
  [\,Q_{\dot{\alpha}},\,[\,R^{[c_1c_2c_3},\,R^{c_4c_5c_6]}\,]\,] 
  &=& [\,[\,Q_{\dot{\alpha}},\,R^{[c_1c_2c_3}\,],\,R^{c_4c_5c_6]}\,]\,] ~+~
      [\,R^{[c_1c_2c_3},\,[\,Q_{\dot{\alpha}},\,R^{c_4c_5c_6]}\,]\,]\\[2ex]
  &=& \delta\kappa\,
      \left(\,
               [\,\Gamma^{[c_1c_2c_3},\,\Gamma^{c_4c_5c_6]}\,]\,
      \right)_{\dot{\alpha}}{}^{\dot{\gamma}}\,Q_{\dot{\gamma}}\\
  &=& 2\,\delta\kappa\,\left(\,\Gamma^{c_1c_2c_3c_4c_5c_6}\,\right)_{\dot{\alpha}}{}^{\dot{\gamma}}\,Q_{\dot{\gamma}}
\end{eqnarray*}
which can be seen as the definition for the action of $Q_{\dot{\alpha}}$ 
on $R^{c_1\ldots c_6}$:
\begin{eqnarray}\label{QR6}
  [\,Q_{\dot{\alpha}},\,R^{c_1\ldots c_6}\,] 
  &=&  \frac{2\,\delta\kappa}{c_{3,3}}\,
       \left(\,
               \Gamma^{c_1c_2c_3c_4c_5c_6}\,
       \right)_{\dot{\alpha}}{}^{\dot{\gamma}}\,Q_{\dot{\gamma}}
\end{eqnarray}
\end{proof}

\begin{proof} of Nr.~(\ref{Jid_24})\\

\noindent 
By the same reasoning as in the proof of Nr.~(\ref{Jid_23}) we find 
\begin{eqnarray*}
  [\,Q_{\dot{\alpha}},\,[\,R^{[c_1\ldots c_6},\,R^{d_1\ldots d_6]}\,]\,] 
  &=& 2\,\left(\,\frac{2\delta\kappa}{c_{3,3}}\,\right)^2\,
      \left(\,
               \Gamma^{[c_1\ldots c_6d_1\ldots d_6]}\,
      \right)_{\dot{\alpha}}{}^{\dot{\gamma}}\,Q_{\dot{\gamma}} 
  ~\buildrel ! \over =~ 0.
\end{eqnarray*}
The rhs vanishes due to the anti symmetrisation of 12 out of 11 
indices. The lhs vanishes due to the bosonic algebra. 
\end{proof}

\begin{proof} of Nr.~(\ref{Jid_28})\\

\noindent 
\begin{eqnarray*}
   [\,Q_{\dot{\alpha}},\,[\,K^c{}_d,\,R^{c_1c_2c_3}\,]\,] &=& 
   [\,[\,Q_{\dot{\alpha}},\,K^c{}_d\,],\,R^{c_1c_2c_3}\,] ~+~
   [\,K^c{}_d,\,[\,Q_{\dot{\alpha}},\,R^{c_1c_2c_3}\,]\,]
\end{eqnarray*}
\begin{eqnarray*}
  {\rm lhs} &=& [\,Q_{\dot{\alpha}},\,
                   \delta_d^{c_1}\,R^{cc_2c_3} ~+~
                   \delta_d^{c_2}\,R^{c_1cc_3} ~+~
                   \delta_d^{c_3}\,R^{c_1c_2c}\,]\\
            &=& \delta\,
                \left\{\,
                   \delta_d^{c_1}\,\Gamma^{cc_2c_3} ~+~
                   \delta_d^{c_2}\,\Gamma^{c_1cc_3} ~+~
                   \delta_d^{c_3}\,\Gamma^{c_1c_2c}\, 
                \right\}_{\dot{\alpha}}{}^{\dot{\beta}}\,
                \tilde{Q}_{\dot{\beta}}\\[2ex]
 {\rm rhs} &=& \delta\,([\,k^c{}_d,\,\Gamma^{c_1c_2c_3}\,])_{\dot{\alpha}}
               {}^{\dot{\beta}}\,\tilde{Q}_{\dot{\beta}}
\end{eqnarray*}
Together\footnote
{ The antisymmetric part of $k^{[cd]}\,=\,\frac{1}{4}\Gamma^{cd}$ gives the 
  Gamma-matrix identity
  $[\,\frac{1}{4}\Gamma^{cd},\,\Gamma_{c_1c_2c_3}\,]
  \,=\,3\cdot \delta^{[d}_{[c_1}\,\Gamma^{c]}{}_{c_2c_3]}$
}
%
\begin{eqnarray*}
   [\,k^c{}_d,\,\Gamma^{c_1c_2c_3}\,] 
   &=& \delta_d^{p_1}\,\Gamma^{cp_2p_3} ~+~
       \delta_d^{p_2}\,\Gamma^{p_1cp_3} ~+~
       \delta_d^{p_3}\,\Gamma^{p_1p_2c}
\end{eqnarray*}

\end{proof}

\begin{proof} of Nr.~(\ref{Jid_29})\\

\noindent 
Completely analogous to Nr.~(\ref{Jid_28}).
\end{proof}

\begin{proof} of Nr.~(\ref{Jid_30})\\

\noindent 
Again analogous to the proof of Nr.~(\ref{Jid_24}) we find 
\begin{eqnarray*}
  [\,Q_{\dot{\alpha}},\,[\,R^{[c_1c_2c_3},\,R^{d_1\ldots d_6]}\,]\,] 
  &=& \frac{2\,\delta^2\kappa}{c_{3,3}}\,
      \left(\,
              [\,\Gamma^{[c_1c_2c_3},\,\Gamma^{d_1\ldots d_6]}\,]\,
      \right)_{\dot{\alpha}}{}^{\dot{\gamma}}\,Q_{\dot{\gamma}} 
  ~\buildrel ! \over =~ 0.
\end{eqnarray*}
\end{proof}

\hskip-1ex
\subsection{Jacobis with two fermionic generators}

\begin{center}
\refstepcounter{table}
\label{TwoFermionJacobianIdentities_Q}
\begin{tabular}{|c|c|}
\hline
    &     \\[-1ex]
 Nr.& l.h.s. of eq.~(\ref{JacobiIdentity})\\[0.5ex]
\hline
    &     \\[-1ex]
    \refstepcounter{nummer}\label{Jid_31}(\thenummer) & 
    $[\,\mathfrak{G},\,[\,Q_{\dot{\alpha}},\,Q_{\dot{\beta}}\,]\,]$ \\
    \refstepcounter{nummer}\label{Jid_32}(\thenummer) & 
    $[\,\mathfrak{G},\,[\,\tilde{Q}_{\dot{\alpha}},\,Q_{\dot{\beta}}\,]\,]$ \\
    \refstepcounter{nummer}\label{Jid_33}(\thenummer) & 
    $[\,\mathfrak{G},\,[\,\tilde{Q}_{\dot{\alpha}},\,\tilde{Q}_{\dot{\beta}}\,]\,]$ \\[1ex]
\hline
\end{tabular}
\center{{\bf Tab.~\thetable~} Two fermionic $Q_{\dot{\alpha}}$-generators}
\end{center}
Defines the action of the bosonic generators
$\mathfrak{G}\,=\,\{\,K^a{}_b,\,R^{c_1c_2c_3},\,R^{c_1\ldots c_6}\,\}$ on the
``central charges'' of the supersymmetry algebras, {\it i.e.} on
$Z_{\dot{\alpha}\dot{\beta}}$, $\tilde{Z}_{\dot{\alpha}\dot{\beta}}$ and 
$A_{\dot{\alpha}\dot{\beta}}$ defined by 
\begin{eqnarray}\label{ZentrLadung}
    \{\,Q_{\dot{\alpha}},\,Q_{\dot{\beta}}\,\} &=& Z_{\dot{\alpha}\dot{\beta}},
    \qquad
    \{\,\tilde{Q}_{\dot{\alpha}},\,\tilde{Q}_{\dot{\beta}}\,\} ~=~
    \tilde{Z}_{\dot{\alpha}\dot{\beta}},\qquad
    \{\,\tilde{Q}_{\dot{\alpha}},\,Q_{\dot{\beta}}\,\} ~=~
    A_{\dot{\alpha}\dot{\beta}}
\end{eqnarray}

\begin{proof} of Nr.~\ref{Jid_31}:\\

\noindent
We expand $Z_{\dot{\alpha}\dot{\beta}}$ in the Clifford algebra 
\begin{eqnarray*}
    Z_{\dot{\alpha}\dot{\beta}} 
    &=& \Gamma^c P_c \,+\, \frac{1}{2!}\,\Gamma^{c_1c_2}Z_{c_1c_2}
                     \,+\, \frac{1}{5!}\,
        \Gamma^{c_1\ldots c_5}Z_{c_1\ldots c_5}  
\end{eqnarray*}
and similar expressions hold for $\tilde{Z}_{\dot{\alpha}\dot{\beta}}$ and 
$A_{\dot{\alpha}\dot{\beta}}$. After repeated application of 
the Jacobi identities and the formula $Tr(\Gamma^{a_i..a_1}\Gamma_{b_1\ldots
  b_j})\,=\,32\cdot \delta^{a_1\ldots a_i}_{b_1\ldots b_j}$ one obtains  
{\it e.g.} for ${\mathfrak{G}}_1\,=\,P_a$:
\begin{eqnarray*}
    [\,R^{c_1c_2c_5},\,(\Gamma^a)^{\dot{\beta}\dot{\alpha}}Z_{\dot{\alpha}\dot{\beta}}\,]
    &=&
    \frac{6}{2!}\,\delta\,\eta^{a[c_1}\delta^{c_2c_3]}_{[b_1b_2]}\cdot
    32\cdot A^{b_1b_2}
\end{eqnarray*}
or finally 
\begin{eqnarray*}
    [\,P_a,\,R^{c_1c_2c_3}\,]
    &=& -\,6\,\delta\,\eta^{a[c_1}A^{c_1c_3]}
\end{eqnarray*}

\end{proof}

\subsection{Jacobis with three fermionic generators}

\begin{center}
\refstepcounter{table}
\label{ThreeFermionJacobianIdentities_Q}
\begin{tabular}{|c|l|c|}
\hline
    &      &\\[-1ex]
 Nr.& l.h.s. of eq.~(\ref{JacobiIdentity}) & satisfied ? \\[0.5ex]
\hline
    &      &\\[-1ex]
    \refstepcounter{nummer}\label{Jid_34}(\thenummer) & 
    $[\,Q_{\dot{\alpha}},\,[\,Q_{\dot{\beta}} ,\,Q_{\dot{\gamma}}\,]\,]$ &
    {\sl cf. proof}\\
    \refstepcounter{nummer}\label{Jid_35}(\thenummer) & 
    $[\,\tilde{Q}_{\dot{\alpha}},\,[\,\tilde{Q}_{\dot{\beta}} ,\,\tilde{Q}_{\dot{\gamma}}\,]\,]$ &
    {\sl cf. proof}\\
    \refstepcounter{nummer}\label{Jid_36}(\thenummer) & 
    $[\,\tilde{Q}_{\dot{\alpha}},\,[\,Q_{\dot{\beta}} ,\,Q_{\dot{\gamma}}\,]\,]$ &
    {\sl cf. proof}\\
    \refstepcounter{nummer}\label{Jid_37}(\thenummer) & 
    $[\,Q_{\dot{\alpha}},\,[\,\tilde{Q}_{\dot{\beta}} ,\,\tilde{Q}_{\dot{\gamma}}\,]\,]$ &
    {\sl cf. proof}\\[1ex]
\hline
\end{tabular}
\center{{\bf Tab.~\thetable~} Three fermionic $Q_{\dot{\alpha}}$-generator}
\end{center}

\begin{proof} of Nr.~(\ref{Jid_34}) and Nr.~(\ref{Jid_35})\\

\noindent
Are just the statements, that $Q_{\dot{\alpha}}$ ($\,\tilde{Q}_{\dot{\alpha}}\,$) act trivial on the generators appearing on the right hand side of the 
$\{Q_{\dot{\beta}},\,Q_{\dot{\gamma}}\}$ ($\,\{\tilde{Q}_{\dot{\beta}},\,\tilde{Q}_{\dot{\gamma}}\}\,$) anticommutator.
\end{proof}
 
\begin{proof} of Nr.~(\ref{Jid_36})\\

\noindent
\begin{eqnarray*}
    [\,\tilde{Q}_{\dot{\alpha}},\,[\,Q_{\dot{\beta}} ,\,Q_{\dot{\gamma}}\,]\,]
&=& [\,[\,\tilde{Q}_{\dot{\alpha}},\,Q_{\dot{\beta}}\,],\,Q_{\dot{\gamma}}\,]
~-~ [\,Q_{\dot{\beta}},\,[\,\tilde{Q}_{\dot{\alpha}}
,\,Q_{\dot{\gamma}}\,]\,]
\end{eqnarray*}

\begin{eqnarray*}
    [\,\tilde{Q}_{\dot{\alpha}},\,Z_{\dot{\beta}\dot{\gamma}}\,]
&=& -2\,[\,Q_{(\dot{\beta}}\,,A_{|\dot{\alpha}|\dot{\gamma})}\,]
\end{eqnarray*}
\end{proof}

\begin{proof} of Nr.~(\ref{Jid_37})\\

\noindent
Completely analogous to proof Nr.~(\ref{Jid_36}) and leads to:
\begin{eqnarray*}
    [\,Q_{\dot{\alpha}},\,\tilde{Z}_{\dot{\beta}\dot{\gamma}}\,]
&=& -2\,[\,\tilde{Q}_{(\dot{\beta}}\,,A_{\dot{\gamma})\dot{\alpha}}\,]
\end{eqnarray*}
\end{proof}

\noindent
\begin{rem}\label{Rem3Q} The action of $Q_{\dot{\alpha}}$ and
$\tilde{Q}_{\dot{\alpha}}$ on $A_{\beta\gamma}$ fixes the action of 
$Q_{\dot{\alpha}}$ and $\tilde{Q}_{\dot{\alpha}}$ on $Z_{\beta\gamma}$ 
and $\tilde{Z}_{\beta\gamma}$. It is consistent to set this action to zero. 
Then all the three algebras $Z_{\beta\gamma}$,
$\tilde{Z}_{\beta\gamma}$ and  $A_{\beta\gamma}$ are abelian and don't mix
with each other. 
\end{rem}

\section{Formulas}

\begin{eqnarray}
     e^{-A}\,d\,e^{A} &=& dA ~-~ \frac{1}{2!}\,[\,A,\,dA\,]
                      ~+~ \frac{1}{3!}\,[\,A,\,[\,A,\,dA\,]\,] 
                      ~+~ \ldots\label{d_Expansion}\\
     e^{-A}\,B\,e^{A} &=& B ~-~ [\,A,\,B\,] 
                      ~+~ \frac{1}{2!}\,[\,A,\,[\,A,\,B\,]\,]
                      ~+~ \ldots\label{B_Expansion}
\end{eqnarray}

\end{appendix}

\end{document}